\documentclass[aps,preprint,showpacs,floats,epsf,epsfig,nofootinbib,12pt]{revtex4}
\usepackage{graphicx}
\usepackage{epsfig}
\usepackage[dvips]{color}
\usepackage{subfigure}

\def\beq{\begin{eqnarray}}
\def\eeq{\end{eqnarray}}

\def\ra{\rangle}

\begin{document}

\title{Study on decays of $Z_c(4020)$ and $Z_c(3900)$  into $h_c+\pi$}

\vspace{1cm}

\author{ Hong-Wei Ke$^1$\footnote{khw020056@hotmail.com} and
        Xue-Qian Li$^2$\footnote{lixq@nankai.edu.cn}  }

\affiliation{  $^{1}$ School of Science, Tianjin University, Tianjin 300072, China \\
  $^{2}$ School of Physics, Nankai University, Tianjin 300071, China }

\vspace{12cm}

\begin{abstract}
At the invariant mass spectrum of  $h_c\pi^\pm$ a
new resonance $Z_c(4020)$ has been observed, however
the previously confirmed $Z_c(3900)$ does not show up at this channel. 
In this paper
we assume that $Z_c(3900)$ and $Z_c(4020)$ are molecular
states of $D\bar D^*(D^{*} \bar D)$ and  $D^*\bar D^*$ respectively,
then we calculate the transition rates of $Z_c(3900)\to
h_c+\pi$ and $Z_c(4020)\to h_c+\pi$ in the light front model. Our results show
that the partial width of $Z_c(3900)\to h_c+\pi$ is only three times smaller
than that of $Z_c(4020)\to h_c+\pi$.
$Z_c(4020)$ seems to be a molecular state, so if $Z_c(3900)$ is also a
molecular state it should be observed in the portal
$e^+e^-\to h_c\pi^\pm$  as long as the database is sufficiently large, by contrary if the future
more precise measurements still cannot find
$Z_c(3900)$ at $h_c\pi^\pm$ channel, the molecular assignment to $Z_c(3900)$  should be ruled out.


\end{abstract}

\maketitle

\section{Introduction}

Since discovery of the exotic XYZ particles and as well as the
pentaquarks, to determine their inner structure and relevant
physics composes a challenge to our understanding of the basic
principles, especially the non-perturbative QCD effects. Gaining
knowledge on their inner structure can only be realized through
analyzing their production and decays behaviors, absolutely, it is
indirect, but efficient. In 2013 the BES collaboration observed a
new resonance $Z_c(4020)$ at the $h_c\pi^{\pm}$ invariant mass
spectrum by studying the process $e^+e^-\to h_c\pi^+\pi^-$ with
the center-of-mass energies from 3.90 GeV to 4.42
GeV\cite{Ablikim:2013wzq}. Its mass and width are $(4022.9\pm
0.8\pm2.7)$MeV and $(7.9\pm 2.7\pm 2.6)$MeV. Recently the neutral
charmonium-like partner of $Z_c(4020)^0$ has also been
experimentally observed \cite{Ablikim:2014dxl}. In 2013
$Z_c(3900)$ was measured at the invariant mass spectrum of
$J/\psi\pi^{\pm}$ with the mass and width being
$(3.899\pm3.6\pm4.9)$ GeV and $(46\pm10\pm10)$ MeV
respectively\cite{Ablikim:2013mio,Liu:2013dau,Xiao:2013iha}. Since
the new resonances $Z_c(4020)$ and $Z_c(3900)$ are charged, they
cannot be charmonia, but their masses and decay modes imply that
they are hidden charm states, namely should be exotic states with
a $c\bar cq\bar q'$ structure. The authors of Ref.
\cite{Esposito:2014hsa,Wang:2014gwa,Deng:2014gqa,Li:2013xia}
considered that the two resonances should be studied in a unique
theoretical framework due to their similarity. It is suggested
that the two resonances could be molecular
states\cite{Wang:2014gwa,Cui:2013yva,Zhang:2013aoa,Wang:2013cya,Wilbring:2013cha},
whereas some other authors regard them as
tetraquark\cite{Deng:2014gqa}, a mixture of the two
structures\cite{Voloshin:2013dpa}\textbf{ or a cusp
structure}\cite{Liu:2013vfa}. The key point is whether one can use
an effective way to confirm their structures. No doubt, it must be
done through combing careful theoretical studies and precise
measurements in the coming experiments.

Even though the masses of the two resonances are close, but their widths
are quite apart, especially  at present
no significant $Z_c(3900)$ signal has been observed at the $h_c\pi^{\pm}$
mass spectrum through the process $e^+e^-\to
h_c\pi^+\pi^-$\cite{Ablikim:2013wzq}. Its absence may imply that
the two resonances might be different, but do we have an evidence to make
a conclusion? If they are of different inner structures, their decay modes should be
different, i.e. different
structures would lead to different decay rates for the same channel which can be tested by
more precise measurements.
Theoretically assigning a special structure to any of $Z_c(3900)$ and $Z_c(4020)$, one can predict its decay rate in an
appointed channel and then the data would tell if the assignment is valid or should be negated. That is the
strategy of this work.

In our early paper.\cite{Ke:2013gia} we explored some strong
decays of $Z_c(3900)$ and $Z_c(4020)$ which were assumed to be
molecular states of $D\bar D^*(D^*\bar D)$ and $D^*\bar D^*$ and the achieved
numerical results are satisfactorily consistent with experimental observations. In this
paper we are going to study the strong decays
$Z_c(3900)\rightarrow h_c\pi$ and $Z_c(4020)\rightarrow h_c\pi$
with the same method.

In order to explore the decays of a molecular state\cite{Ke:2013gia}, we extended the light front quark model
(LFQM) which was thoroughly studied in literature
\cite{Jaus,Ji:1992yf,Cheng:2004cc,Cheng:1996if,Cheng:2003sm,Choi:2007se,
Hwang:2006cua,Ke:2007tg,Ke:2009ed,Li:2010bb,Ke:2013zs}. In this situation
the constituents are two mesons instead of a quark and an
antiquark in the light front frame. In the case of covariant form the
constituents are off-shell. The effective interactions between the
two constituent mesons are adopted following the literature
\cite{Haglin:1999xs,Oh:2000qr,Lin:1999ad,Deandrea:2003pv,Meng:2007cx,Yuan:2012zw},
where by fitting relevant processes, the effective coupling constants have
been obtained.  Using the method given in
Ref.\cite{Ke:2013gia}  we deduce the corresponding form
factors and estimate the decay widths of
$Z_c(3900)\rightarrow h_c\pi$ and $Z_c(4020)\rightarrow h_c\pi$ while
both $Z_c(3900)$ and $Z_c(4020)$ are assumed to be molecular states. In
fact there exist three degenerate S-wave bound states of
$D^*\bar D^*$ whose quantum numbers are respectively $0^+$, $1^+$ and $2^+$.
In our work we evaluate the decay rates of the $D^*\bar
D^*$ molecules which can be either of the three quantum states.

In this framework, the  $q^+=0$ condition is applied i.e. $q^2<0$,
it means that the final mesons are not on-shell, thus the
obtained form factors are space-like. Then one
needs to extrapolate analytically the form factors from the un-physical space-like
region to the time-like region to reach the physical ones.  With
the form factors we calculate the corresponding decay widths. The
numerical results will provide us with information about the structures of
$Z_c(3900)$ and $Z_c(4020)$.

After the introduction we derive the form factors for transitions
$Z_c(3900)\rightarrow h_c\pi$ and $Z_c(4020)\rightarrow h_c \pi$
in section II. Then we numerically evaluate the relevant form
factors and decay widths in section III. In the last section we discuss  the
numerical results and draw our conclusion. Some details about the
approach are collected in the Appendix.

\section{the strong decays  $Z_c(3900)\rightarrow h_c+\pi$}
In this section we calculate the strong decay rate of $Z_c(3900)\rightarrow
h_c+\pi$, while assuming $Z_c(3900)$ as a $1^{+}$ $D\bar D^*$ molecular
state, in the light-front model. Since the success of applying the method \cite{Ke:2013gia} we
we have reason to believe that this framework also works in this case. The
configuration of the $D\bar D^*$ molecular state is
$\frac{1}{\sqrt{2}}(D\bar D^*+\bar D D^*)$.
\begin{figure}
        \centering
        \subfigure[~]{
          \includegraphics[width=8cm]{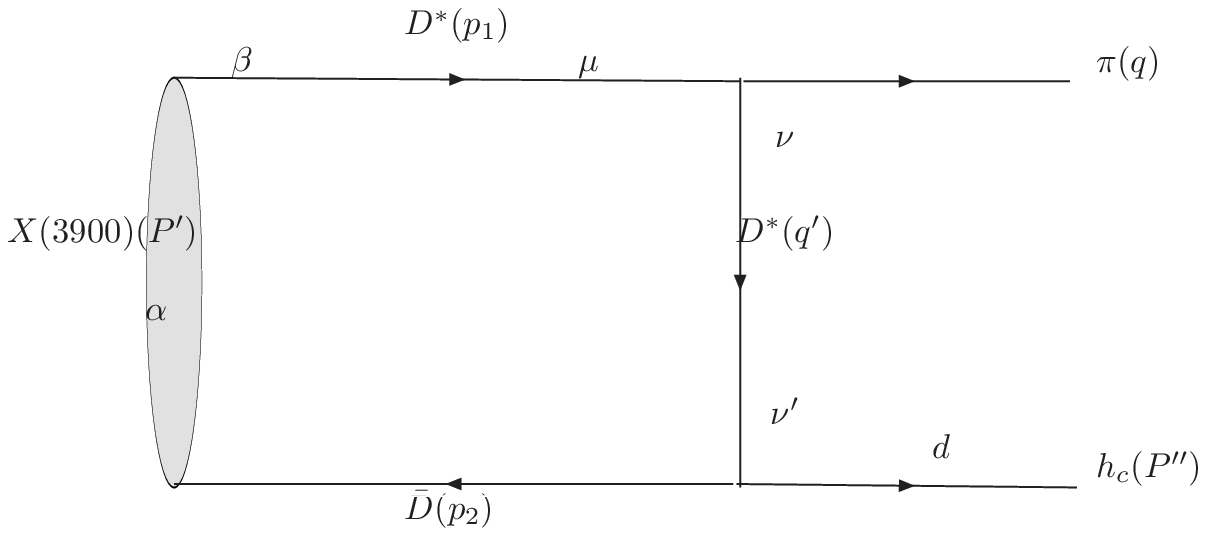}}
        \subfigure[~]{
          \includegraphics[width=8cm]{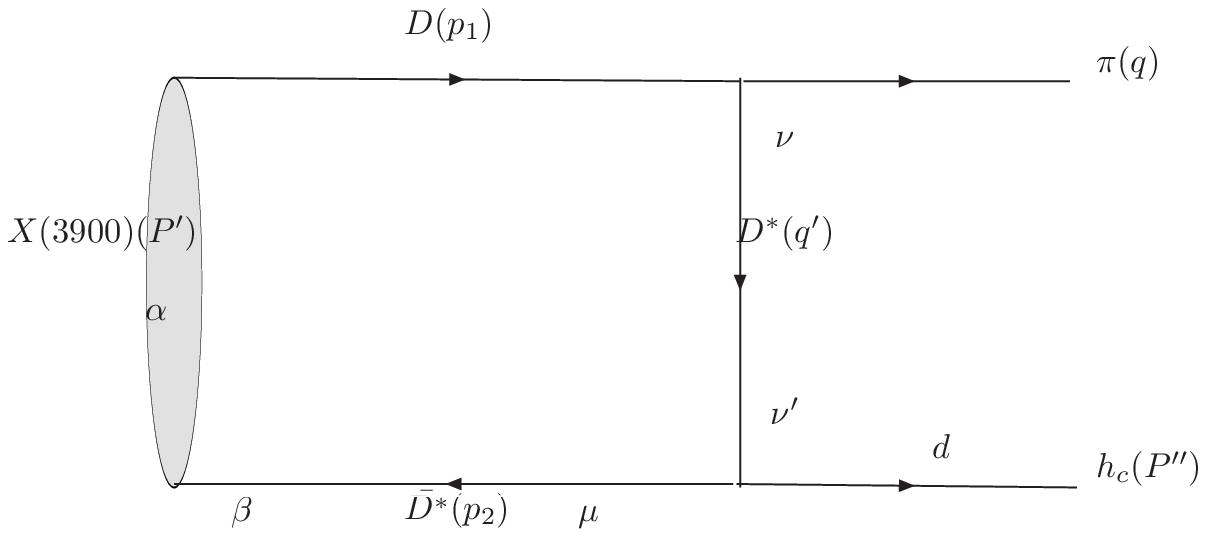}}
\caption{Strong decays of molecular states ( two diagrams where
$h_c$ and $\pi$ in the final states are switched are omitted).}
        \label{fig1}
    \end{figure}
The Feynman diagrams for $Z_c(3900)$ decaying into $h_c\pi$ by
exchanging $D$ or $D^*$  mesons are shown in Fig.\ref{fig1}.

Following Ref.\cite{Cheng:2003sm}, the hadronic matrix
element corresponding to the diagrams  in Fig.\ref{fig1} is written as
\begin{eqnarray}
{\mathcal{A}}_{11}=i\frac{1}{(2\pi)^4}\int d^4
p_1\frac{[H_{A_{01}}S^{1(a)}_{d\alpha}+H_{A_{10}}S^{1(b)}_{d\alpha}]}{N_1N_1'N_2}\epsilon_1^d\epsilon^\alpha
\end{eqnarray}\label{eq1}
with
\begin{eqnarray} S_{d\alpha}^{1(a)}&&=-i\frac{g_{_{h_c
D^*D^*}}g_{_{\pi
DD^*}}}{\sqrt{2}}g_{\alpha\beta}g^{\mu\beta}g^{\nu\nu'}\varepsilon_{a\mu
c\nu}p_1^a (p_1^{c}-q^{c})g_{d\nu'}
\mathcal{F}(m_1,p_1)\mathcal{F}(m_2,p_2)\mathcal{F}^2(m_{D^*},q'),
\nonumber\\S_{d\alpha}^{1(b)}&&=i\frac{g_{_{h_c DD^*}}g_{_{\pi
D^*D^*}}}{\sqrt{2}}g_{\alpha\beta}g^{\mu\beta}g^{\nu\nu'}(p_{1\nu}+
q_{\nu})P''^\omega \varepsilon_{\omega
d\mu\nu'}\mathcal{F}(m_1,p_1)\mathcal{F}(m_2,p_2)\mathcal{F}^2(m_{D^*},q'),
\end{eqnarray} $N_1=p_1^2-m_1^2+i\varepsilon$,
where $N_1'={q'}^2-m_{q'}^2+i\varepsilon$ and
$N_2=p_2^2-m_2^2+i\varepsilon$. A form factor
$\mathcal{F}(m_i,p^2)=\frac{(m_i+\Lambda)^2-m_i^2}{(m_i+\Lambda)^2-p^2}$
is introduced to compensates the off-shell effect caused by the
intermediate meson of mass $m_i$ and momentum $p$.  $H_{A_{10}}$
and $H_{A_{01}}$ are vertex functions which include the normalized
wavefunctions of the decaying mesons with the assigned quantum
numbers and are invariant in the four-dimension. In fact, for the
practical computation their exact forms are not necessary, because
after integrating over $dp_1^-$ the integral is reduced into a
three-dimensional integration, and $H_{A_{10}}$( or $H_{A_{01}}$)
would be replaced by $h_{A_{10}}$ ( $h_{A_{01}}$) whose explicit
form(s) is calculable. In the light-front frame the momentum $p_i$
is decomposed into its components as ($p_i^-,p_i^+,{p_i}_\perp$)
and integrating out $p_{1}^-$ with the methods given in
Ref.\cite{Cheng:1996if} one has
\begin{eqnarray}\label{vf9.2}
\int d^4p_1 \frac{H_{A}S_{d\alpha}}{N_1N_1'N_2}{\epsilon_1}^d
{\epsilon}^{\alpha}\rightarrow-i\pi\int
dx_1d^2p_\perp\frac{h_{A}\hat S_{d\alpha}}{x_2
\hat{N_1}\hat{N_1'}} {\epsilon_1}^d {\epsilon}^{\alpha},
\end{eqnarray}
with
\begin{eqnarray*}
&&\hat{N}_1=x_1({M}^2-{{M}_0}^2),\\
&&\hat{N}_1^{'}=x_2q^2-x_1{{M}_0}^2+x_1M'^2+2p_\perp\cdot
q_\perp,\\&&h_{A}=\sqrt{\frac{x_1x_2}{m_1m_2}}(M^2-{M}_0^2)h_{A}'
\end{eqnarray*}
where $M$ and $M'$ are the masses of initial and finial mesons.
The factor $\sqrt{x_1x_2}(M^2-{M}_0^2)$ in the expression of
$h_{A}$ was introduced \cite{Cheng:2003sm} and an additional
normalization factor $\sqrt{\frac{1}{m_1m_2}}$ appears
corresponding to the boson constituents in the molecular state.
The explicit expressions of the effective form factors $h_{A}'$
are collected in the Appendix.

{Since we calculate the transition in the $q^+=0$ frame the zero
mode contributions  which come from the residues of virtual pair
creation processes, are not involved. To include the
contributions, ${p_1}_\mu$, ${p_1}_\nu$ and ${p_1}_\mu {p_1}_\nu$
in $s_{\mu\nu}^a$ must be replaced by the appropriate expressions
as discussed in Ref.\cite{Cheng:2003sm}}
\begin{eqnarray}\label{eq2}
&&{p_1}_\mu\rightarrow\mathcal{P}_\mu
A^{(1)}_1+q_\mu A^{(1)}_2,\nonumber\\
&&{p_1}_\mu{p_1}_\nu\rightarrow
g_{\mu\nu}A_1^{(2)}+\mathcal{P}_\mu\mathcal{P}_\nu
A_2^{(2)}+(\mathcal{P}_\mu q_\nu+q_\mu\mathcal{P}_\nu)
A_3^{(2)}+q_\mu q_\nu A_4^{(2)}
\end{eqnarray}
where $\mathcal{P}=P'+P''$ and  $q=P'-P''$ with $P'$ and $P''$
denote the momenta of the concerned mesons in the initial and
final states respectively.

For example, after the replacement $S_{d\alpha}^{1(a)}$ turns into
\begin{eqnarray} \hat{S}_{d\alpha}^{1(a)}&&=-i\frac{g_{_{h_c
D^*D^*}}g_{_{\pi
DD^*}}}{\sqrt{2}}g_{\alpha\beta}g^{\mu\beta}g^{\nu\nu'}\varepsilon_{a\mu
c\nu} [g^{ac}A_1^{(2)}+\mathcal{P}^a\mathcal{P}^c
A_2^{(2)}+(\mathcal{P}^a q^c+q^a\mathcal{P}^c) A_3^{(2)}+q^a q^c
A_4^{(2)}\nonumber\\&&\,\,\,\,\,\,\,-(\mathcal{P}^a A^{(1)}_1+q^a
A^{(1)}_2)q^{c}]g_{d\nu'}
\mathcal{F}(m_1,p_1)\mathcal{F}(m_2,p_2)\mathcal{F}^2(m_{D^*},q'),
\nonumber\\&&=i\frac{g_{_{h_c D^*D^*}}g_{_{\pi
DD^*}}}{\sqrt{2}}(A^{(1)}_1- A_3^{(2)}
)\mathcal{P}_aq_c\varepsilon_{acd\alpha}
\mathcal{F}(m_1,p_1)\mathcal{F}(m_2,p_2)\mathcal{F}^2(m_{D^*},q')\nonumber\\&&=i\frac{g_{_{h_c
D^*D^*}}g_{_{\pi DD^*}}}{\sqrt{2}}2(A^{(1)}_1- A_3^{(2)}
){P_a'}q_b\varepsilon_{abd\alpha}
\mathcal{F}(m_1,p_1)\mathcal{F}(m_2,p_2)\mathcal{F}^2(m_{D^*},q'),
\end{eqnarray}

Some notations such as $A_i^{(j)}$ and $M_0'$  can be found in
Ref.\cite{Cheng:2003sm}. With the replacement, $h_A\hat
S_{d\alpha}$ is decomposed into
\begin{eqnarray}\label{eq3}i{F_{1}}P'_aq_b
\varepsilon_{abd\alpha},
\end{eqnarray}
with
\begin{eqnarray}
{F_{1}}=&&{\sqrt{2}}{g_{_{h_c D^*D^*}}g_{_{\pi DD^*}}
h_{A_{01}}}\,\left({A_1^{(1)}} - {A_3^{(2)}} \right)
\mathcal{F}(m_1,p_1)\mathcal{F}(m_2,p_2)\mathcal{F}^2(m_{D^*},q')\nonumber\\&&+\frac{g_{_{h_c
DD}}g_{_{\pi DD^*}}}{\sqrt{2}} h_{A_{10}}\left({A_1^{(1)}} +
{A_2^{(1)}}
+1\right)\,\mathcal{F}(m_1,p_1)\mathcal{F}(m_2,p_2)\mathcal{F}^2(m_{D},q').
\end{eqnarray}\label{eq5}

For convenience of derivation, let us introduce a new form factor which is defined as following
\begin{eqnarray}\label{eq60}
&&{f_{1}}(m_1,m_2)=\frac{1}{16\pi^3}\int
dx_2d^2p_\perp\frac{F_{1}}{x_2 \hat{N_1}\hat{N_1'}}.
\end{eqnarray}
Then the amplitude is written in terms of $f_{1}(m_1,m_2)$ as
\begin{eqnarray}\label{eq7}
{\mathcal{A}}_{11}&&= i{f_{1}}(m_1,m_2)P'_aq_b
\varepsilon_{abd\alpha}\epsilon_1^d\epsilon^\alpha.
\end{eqnarray}

The contributions from the Feynman diagrams by switching around
$h_c$ and $\pi$ in the final states (in  Fig.1) can be formulated by simply
exchanging $m_1$ and $m_2$ in the expression ${f_{1}}(m_1,m_2)$.
Then the total amplitude is
\begin{eqnarray}\label{eq8}
\mathcal{A}_1=i[f_{1}(m_1,m_2)+f_{1}(m_2,m_1)] P'_aq_b
\varepsilon_{abd\alpha}=ig_1 P'_aq_b
\varepsilon_{abd\alpha}\epsilon_1^d\epsilon^\alpha,
\end{eqnarray}
and the factor $g_1$ will be numerically evaluated in next section.

\section{The strong decay  $Z_c(4020)\rightarrow h_c+\pi$}
\begin{figure}
        \centering
        \subfigure[~]{
          \includegraphics[width=8cm]{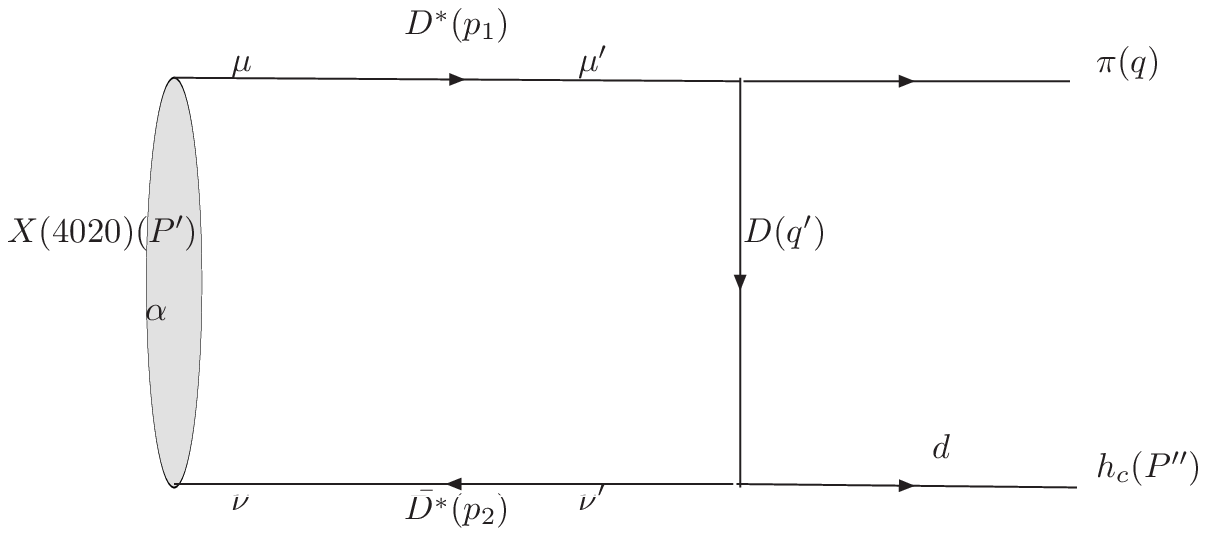}}
        \subfigure[~]{
          \includegraphics[width=8cm]{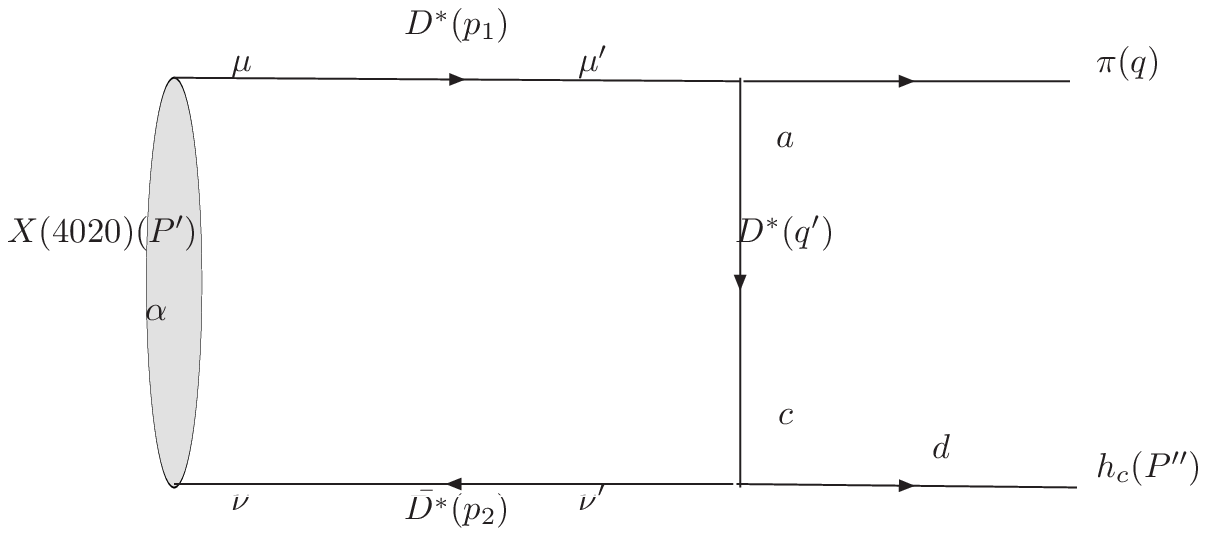}}
\caption{Strong decays $Z_c(4020)\rightarrow h_c\pi $ ( the
Figures exchanged the final states are omitted).}
        \label{Fig3}
    \end{figure}

Similar to what we have
done for $Z_c(3900)$, we calculate the decay rate of
$Z_c(4020)\rightarrow h_c\pi$ by respectively supposing $Z_c(4020)$ as  $0^+$,
$1^+$ and $2^+$ $D^*\bar D^*$ molecular states. The Feynman
diagrams are shown in Fig.\ref{Fig3}.

\subsection{$Z_c(4020)$ as a $0^+$ molecular state }
In terms of the vertex function given in the appendix, the
hadronic matrix element is
\begin{eqnarray}
\mathcal{A}_{21}=i\frac{1}{(2\pi)^4}\int d^4
p_1\frac{H_{A_0}}{N_1N_1'N_2}(S^{2(a)}_{d}+S^{2(b)}_{d})\epsilon_1^d,
\end{eqnarray}\label{eq13}
where
$$S_{d}^{2(a)}=ig_{_{h_c DD^*}}g_{_{\pi DD^*}}g_{\mu\nu}
g^{\mu\mu'}(2q_{\mu'}-p_{1\mu'})g^{\nu\nu'}g_{\nu'd}
\mathcal{F}(m_1,p_1)\mathcal{F}(m_2,p_2)\mathcal{F}^2(m_{D},q'),$$
and $S_{d}^{2(b)}=-ig_{_{h_c D^*D^*}}g_{_{\pi
D^*D^*}}g_{\mu\nu}g^{\mu\mu'}g^{\nu\nu'}\varepsilon_{\omega\mu'\rho
a}
p_1^{\omega}q'^{\rho}g^{ac}P''^f\varepsilon_{fdc\nu}\mathcal{F}(m_1,p_1)\mathcal{F}(m_2,p_2)\mathcal{F}^2(m_{D^*},q').$
Carrying out the integration and making the required replacements,
we have
\begin{eqnarray}\label{eq14}
h_{A_0}(\hat S^{2(a)}_{d}+\hat S^{2(b)}_{d})=i{F_{2}} q_{d},
\end{eqnarray}
with
\begin{eqnarray}\label{eq15}
{F_{2}}=&&{g_{_{\psi DD^*}}g_{_{\pi DD^*}}h_{A_0}}
      \left(2-{A_1^{(1)}} - {A_2^{(1)}} \right)
    \mathcal{F}(m_1,p_1)\mathcal{F}(m_2,p_2)\mathcal{F}^2(m_{D},q')
\nonumber\\&&-{4\,g_{_{\psi D^*D^*}}g_{_{\pi
D^*D^*}}h_{A_0}}\left({A_1^{(1)}} + {A_3^{(2)}}
\right)M''^2\mathcal{F}(m_1,p_1)\mathcal{F}(m_2,p_2)\mathcal{F}^2(m_{D^*},q').
\end{eqnarray}

Simultaneously, we have derived the form factor
\begin{eqnarray}\label{eq16}
&&{f_{2}}(m_1,m_2)=\frac{1}{16\pi^3}\int
dx_2d^2p_\perp\frac{F_{2}}{x_2 \hat{N_1}\hat{N_1'}}.
\end{eqnarray}
With this form factor   the transition amplitude is obtained as
\begin{eqnarray}\label{eq17}
\mathcal{A}_{21}&&= i{f_{2}}(m_1,m_2) q\cdot\epsilon_1.
\end{eqnarray}
Similarly, the amplitude corresponding the Feynman diagrams
where the mesons in the final state are switched around, can be
easily obtained by exchanging $m_1$ and $m_2$. The total amplitude
is
\begin{eqnarray}\label{eq18}
\mathcal{A}_2&&=i[f_{2}(m_1,m_2)+f_{2}(m_2,m_1)]
q\cdot\epsilon_1\nonumber\\&&=ig_{2} q\cdot\epsilon_1.
\end{eqnarray}

\subsection{$Z_c(4020)$ as a $1^+$ molecular state }
For the $1^+$ state, the hadronic matrix element would be
different from the case where $Z_c(4020)$ is assumed to be a $0^+$
meson. Now the hadronic matrix element is written as
\begin{eqnarray}
\mathcal{A}_{31}=i\frac{1}{(2\pi)^4}\int d^4
p_1\frac{H_{A_1}}{N_1N_1'N_2}(S^{2(a)}_{d\alpha}+S^{2(b)}_{d\alpha})\epsilon_1^d\epsilon^\alpha,
\end{eqnarray}\label{eq13}
where
$$S_{d\alpha}^{2(a)}=ig_{_{h_c DD^*}}g_{_{\pi DD^*}}\varepsilon_{\mu\nu\alpha\beta}
g^{\mu\mu'}(2q_{\mu'}-p_{1\mu'})P'^{\beta}g^{\nu\nu'}g_{\nu'd}
\mathcal{F}(m_1,p_1)\mathcal{F}(m_2,p_2)\mathcal{F}^2(m_{D},q'),$$
and
$$S_{d\alpha}^{2(b)}=-ig_{_{h_c D^*D^*}}g_{_{\pi
D^*D^*}}\varepsilon_{\mu\nu\alpha\beta}g^{\mu\mu'}g^{\nu\nu'}P'^{\beta}\varepsilon_{\omega\mu'\rho
a}
p_1^{\omega}q'^{\rho}g^{ac}P''^f\varepsilon_{fdc\nu}\mathcal{F}(m_1,p_1)\mathcal{F}(m_2,p_2)\mathcal{F}^2(m_{D^*},q').$$
After integrating over the momentum, we
have
\begin{eqnarray}\label{eq14}
h_{A_1}(\hat S^{2(a)}_{d\alpha}+\hat
S^{2(b)}_{d\alpha})=i{F_{3}}P'_aq_b \varepsilon_{abd\alpha},
\end{eqnarray}
with
\begin{eqnarray}\label{eq15}
{F_{3}}=&&{g_{_{h_c DD^*}}g_{_{\pi DD^*}}h_{A_1}}
      \left({A_2^{(1)}} - {A_1^{(1)}}-2 \right)
    \mathcal{F}(m_1,p_1)\mathcal{F}(m_2,p_2)\mathcal{F}^2(m_{D},q')
\nonumber\\&&+{g_{_{h_c D^*D^*}}g_{_{\pi
D^*D^*}}h_{A_1}}\left({A_1^{(1)}}+ {A_3^{(2)}}
\right){(M'^2+M''^2-q^2)}\mathcal{F}(m_1,p_1)\mathcal{F}(m_2,p_2)\mathcal{F}^2(m_{D^*},q').
\end{eqnarray}

The form factor  is
\begin{eqnarray}\label{eq16}
&&{f_{3}}(m_1,m_2)=\frac{1}{16\pi^3}\int
dx_2d^2p_\perp\frac{F_{3}}{x_2 \hat{N_1}\hat{N_1'}},
\end{eqnarray}
which will be numerically evaluated.
With these form factors   the transition amplitude is obtained as
\begin{eqnarray}\label{eq17}
\mathcal{A}_{31}&&= i{f_{3}}(m_1,m_2)P'_aq_b
\varepsilon_{abd\alpha}\epsilon_1^d\epsilon^\alpha.
\end{eqnarray}

Including the contributions of the Feynman diagrams where we
switch around $h_c$ and $\pi$ in the final states, the amplitude is
\begin{eqnarray}\label{eq18}
\mathcal{A}_3&&=i[f_{3}(m_1,m_2)+f_{3}(m_2,m_1)] P'_aq_b
\varepsilon_{abd\alpha}\nonumber\\&&=ig_{3} P'_aq_b
\varepsilon_{abd\alpha}\epsilon_1^d\epsilon^\alpha.
\end{eqnarray}

\subsection{$Z_c(4020)$ as a $2^+$ molecular state }
Then as we suppose $Z_c(4020)$ is a $2^+$ molecule, the hadronic
matrix element is
\begin{eqnarray}
\mathcal{A}_{41}=i\frac{1}{(2\pi)^4}\int d^4
p_1\frac{H_{A_1}}{N_1N_1'N_2}(S^{2(a)}_{d\mu\nu}+S^{2(b)}_{d\mu\nu})\epsilon_1^d\epsilon^{\mu\nu},
\end{eqnarray}\label{eq13}
where
$$S_{d\alpha}^{2(a)}=ig_{_{h_c DD^*}}g_{_{\pi DD^*}}
g^{\mu\mu'}(2q_{\mu'}-p_{1\mu'})g^{\nu\nu'}g_{\nu'd}
\mathcal{F}(m_1,p_1)\mathcal{F}(m_2,p_2)\mathcal{F}^2(m_{D},q'),$$
and $S_{d\alpha}^{2(b)}=-ig_{_{h_c D^*D^*}}g_{_{\pi
D^*D^*}}g^{\mu\mu'}g^{\nu\nu'}\varepsilon_{\omega\mu'\rho a}
p_1^{\omega}q'^{\rho}g^{ac}P''^f\varepsilon_{fdc\nu}\mathcal{F}(m_1,p_1)\mathcal{F}(m_2,p_2)\mathcal{F}^2(m_{D^*},q').$
Carrying out the integration, one has
\begin{eqnarray}\label{eq14}
h_{A_1}(\hat S^{2(a)}_{d\alpha}+\hat S^{2(b)}_{d\alpha})=i({F_{4}}
q_{\mu}g_{d\nu}+{F_{5}}q_{\nu}g_{d\mu}+{F_{6}}q_{\nu}q_{d}q_{\mu}),
\end{eqnarray}
with
\begin{eqnarray}\label{eq15}
&&{F_{4}}={g_{_{h_c DD^*}}g_{_{\pi DD^*}}h_{A_1}}
       \left(2+{A_1^{(1)}} - {A_2^{(1)}} \right)
    \mathcal{F}(m_1,p_1)\mathcal{F}(m_2,p_2)\mathcal{F}^2(m_{D},q')
\nonumber\\
&&{F_{5}}=2g_{_{h_c D^*D^*}}g_{_{\pi
D^*D^*}}h_{A_1}\left({A_1^{(1)}} +{A_3^{(2)}}
\right)\frac{(M'^2+M''^2-q^2)}{2}
  \mathcal{F}(m_1,p_1)\mathcal{F}(m_2,p_2)\mathcal{F}^2(m_{D^*},q')\nonumber\\
&&{F_{6}}=2g_{_{h_c D^*D^*}}g_{_{\pi
D^*D^*}}h_{A_1}\left({A_1^{(1)}} +{A_3^{(2)}} \right)
  \mathcal{F}(m_1,p_1)\mathcal{F}(m_2,p_2)\mathcal{F}^2(m_{D^*},q').
\end{eqnarray}

The new form factors are defined as following
\begin{eqnarray}\label{eq16}
&&{f_{a}}(m_1,m_2)=\frac{1}{16\pi^3}\int
dx_2d^2p_\perp\frac{F_{a}}{x_2 \hat{N_1}\hat{N_1'}},
\end{eqnarray}
where the subscript $a$ denotes 4,5 and 6.
Substituting these form factors into the formulae, the transition amplitude is obtained as
\begin{eqnarray}\label{eq17}
\mathcal{A}_{41}&&= i[{f_{4}}(m_1,m_2)
q_{\mu}g_{d\nu}+{f_{5}}(m_1,m_2)
q_{\nu}g_{d\mu}+{f_{6}}(m_1,m_2)q_{\nu}q_{d}q_{\mu})]\epsilon_1^d\epsilon^{\mu\nu}.
\end{eqnarray}

Similarly, as all the contributions  are incorporated, the total amplitude is
\begin{eqnarray}\label{eq18}
\mathcal{A}_4&&=i\{[{f_{4}}(m_1,m_2)+{f_{4}}(m_2,m_1)]
q_{\mu}g_{d\nu}+[{f_{5}}(m_1,m_2)+{f_{5}}(m_2,m_1)]
q_{\nu}g_{d\mu}\nonumber\\&&+[{f_{6}(m_1,m_2)}q_{\nu}q_{d}q_{\mu}+{f_{6}(m_2,m_1)}q_{\nu}q_{d}q_{\mu}]\}\epsilon_1^d\epsilon^{\mu\nu}\nonumber\\&&=
i[g_{4} q_{\mu}g_{d\nu}+g_{5}
q_{\nu}g_{d\mu}+g_{6}q_{\nu}q_{d}q_{\mu}]\epsilon_1^d\epsilon^{\mu\nu}.
\end{eqnarray}

\section{numerical results}
In this section we present our predictions on the decay rates of
$Z_c(3900)\rightarrow h_c\pi$ and $Z_c(4020)\rightarrow h_c\pi$ along with all the input parameters.
First we need to calculate the corresponding form factors which we deduced
in last section. Those formulas involve some parameters which need
to be priori fixed. We use 3.899 GeV\cite{Ablikim:2013mio} as
the mass of $Z_c(3900)$ and the mass of  $Z_c(4020)$ is determined
to be 4.02 GeV. The masses of the involved mesons are set as
$m_{h_c}=3.525$ GeV, $m_{\pi}=0.139$ GeV, $m_{D}=1.869$ GeV and
$m_{D^*}=2.007$ GeV according to the data book\cite{PDG12}. The coupling
constants $g_{_{\pi DD^*}}=8.8$ and $g_{_{\pi D^*D^*}}=9.08$  GeV$^{-1}$
are adopted according to Refs.\cite{Haglin:1999xs,Oh:2000qr}. At present one cannot
fix the couplings  $h_c DD^{*}$ and $h_c D^{*}D^{*}$ from data yet.
However there exists a simple, but approximate relation
$m_Dg_{_{h_c DD^*}}=g_{_{h_c D^*D^*}}$ which is in analog to the case about the couplings
$\psi D^{(*)}D^{(*)}$\cite{Deandrea:2003pv,Meng:2007cx}, so only
one undetermined parameter remains. Since the values of the most
coupling constants are of order $O(1)$, we set $g_{_{h_c D^*D^*}}=1$ as a reasonable
choice. If one could fix $g_{_{h_c D^*D^*}}$ later, he just needs to multiply a
number to the corresponding form factor and it does not affect our final conclusion. The cutoff parameter
$\Lambda$ in the vertex $\mathcal{F}$ was suggested to be set as 0.88
GeV to 1.1 GeV \cite{Meng:2007cx}. In our calculation we
use 0.88 GeV and 1.1 GeV respectively to study the effect on
the results. The parameter $\beta$ in the wavefunction is not very
certain until now. In Ref.\cite{Ke:2013gia} we estimated its value and decided that it
is close to or slightly smaller than  0.631 GeV$^{-1}$ \cite{Ke:2011jf}, and it is the
$\beta$ number for the wavefunction of $J/\psi$.

Since the form factors are derived in the reference frame of $q^+=0$ (
$q^2<0$) i.e. in the space-like region, we need to extend them
to the time-like region by means of the normal procedure
provided in literatures. In Ref.\cite{Cheng:2003sm} a
three-parameter form factor was suggested as
\begin{eqnarray}\label{eq23}
g(q^2)=\frac{g(0)}{
  \left[1-a\left(\frac{q^2}{M_{Z_c}^2}\right)
  +b\left(\frac{q^2}{M_{Z_c}^2}\right)^2\right]}.
 \end{eqnarray}

\begin{table}[!h]
\caption{The  three-parameter form factors with
($\Lambda=0.88$ GeV, $\beta=0.631$ GeV$^{-1}$).}\label{Tab:t1}
\begin{ruledtabular}
\begin{tabular}{cccc}
  $g$    & $g(0)$
& $a$  &  $b$\\\hline
  $g_{1}$  &   -0.253      &   2.72    &  4.60 \\
  $g_{2}$  &   0.364    &  2.75   &  4.70\\
$g_{3}$  &  -0.129     &   2.74    & 3.25 \\
  $g_{4}$  &  -0.243      &   3.24    &7.01 \\
    $g_{5}$  &   -0.486     &   2.41   &2.42 \\
     $g_{6}$  &  -0.0341     &   2.82  &4.88 \\
\end{tabular}
\end{ruledtabular}
\end{table}


\begin{table}[!h]
\caption{The decay widths of some modes ( $\beta=0.631$
GeV$^{-1}$).}\label{Tab:t4}
\begin{ruledtabular}
\begin{tabular}{cc|cc}
  decay mode($\Lambda=0.88$ GeV)   &  width(GeV)&  decay mode($\Lambda=1.1$ GeV)   &  width(GeV)\\\hline
  $Z_c(3900)\rightarrow h_c\pi$  & $5.85\times 10^{-5}$   &  $Z_c(3900)\rightarrow  h_c\pi$  &  $8.91\times 10^{-5}$    \\
   $Z_c(4020)(0^+)\rightarrow  h_c\pi$  & $1.49\times 10^{-4}$&  $Z_c(4020)(0^+)\rightarrow  h_c\pi$  & $2.36\times 10^{-4}$   \\
   $Z_c(4020)(1^+)\rightarrow  h_c\pi$  & $1.51\times 10^{-4}$  & $Z_c(4020)(1^+)\rightarrow  h_c\pi$  & $2.34\times 10^{-4}$
  \\
  $Z_c(4020)(2^+)\rightarrow  h_c\pi$  & $1.54\times 10^{-4}$  & $Z_c(4020)(2^+)\rightarrow  h_c\pi$  &$2.38\times 10^{-4}$

\end{tabular}
\end{ruledtabular}
\end{table}

\begin{figure}
\begin{center}
\begin{tabular}{ccc}
\scalebox{0.8}{\includegraphics{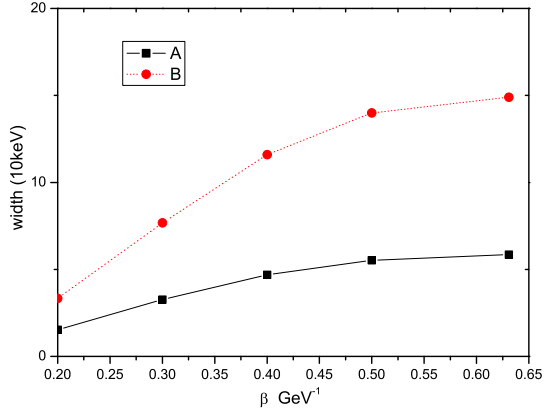}}
\end{tabular}
\end{center}
\caption{the dependence of $\Gamma(Z_c(3900)\rightarrow h_c \pi$)
(A) and $\Gamma(Z_c(4020)\rightarrow h_c \pi$)  (B) on
$\beta$.}\label{Fig6}
\end{figure}

\begin{figure}
\begin{center}
\begin{tabular}{ccc}
\scalebox{0.8}{\includegraphics{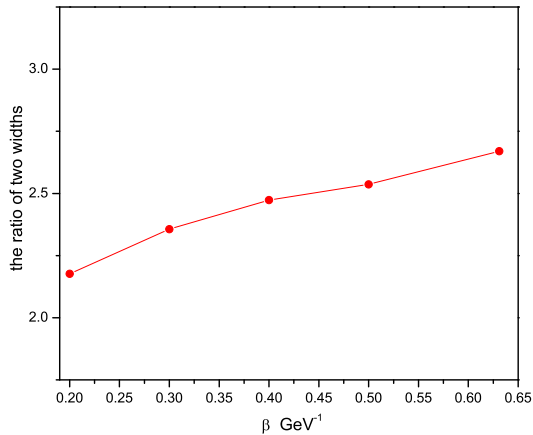}}
\end{tabular}
\end{center}
\caption{the dependence of the ratio $\Gamma(Z_c(4020)\rightarrow
h_c \pi)/\Gamma(Z_c(3900)\rightarrow h_c \pi$) on
$\beta$.}\label{Fig7}
\end{figure}
The resultant form factors are listed in table \ref{Tab:t1} and
the corresponding decay widths are presented in table II. The
molecular states of $D^*\bar D^*$ can be in three different quantum states, thus the
Lorentz structures of their decay amplitudes for $Z_c\to h_c \pi$ are different and
the values of the corresponding form factors should also be different.
However we find that the decay widths of all those states are very close to each other,
and it is easy to understand because the three states with different spin assignments
are degenerate. One can also note,
$\Gamma(Z_c(4020)\rightarrow h_c \pi)$
is three times larger than $\Gamma(Z_c(3900)\rightarrow h_c \pi$)
for different parameter $\Lambda$.

In our calculation, we notice that the model parameter $\beta$ can
affect the numerical results to a certain degree. We illustrate
the dependence of $\Gamma(Z_c(3900)\rightarrow  h_c\pi)$ and
$\Gamma(Z_c(4020)\rightarrow  h_c\pi)$ on $\beta$ in
Fig.\ref{Fig6} and
 depict the dependence of the ratio of $\Gamma(Z_c(4020)\rightarrow  h_c\pi)/\Gamma(Z_c(3900)\rightarrow  h_c\pi)$
  on $\beta$ in  Fig.\ref{Fig7}.
Lines A and B in Fig.\ref{Fig6} correspond to $Z_c(3900)$ and
$Z_c(4020)$ respectively. It is also noted that the ratio
$\Gamma(Z_c(4020)\rightarrow h_c \pi) /\Gamma(Z_c(3900)\rightarrow
h_c \pi)\approx$2.5  does not vary much as $\beta$ changes.

\section{conclusion and discussions}
In this work, supposing  $Z_c(3900)$ and $Z_c(4020)$ to be $D\bar D^*$ and $D^*\bar
D^*$ molecular states,
we calculate the decay rates of $Z_c(3900)\rightarrow  h_c\pi$ and
$Z_c(4020)\rightarrow  h_c\pi$ respectively in the light front model.
It is noted that for the $D^*\bar D^*$
system there are three degenerate states whose quantum numbers are
$0^+$, $1^+$ and $2^+$ with orbital angular momentum $L=0$. Thus we calculate
the decay rates of the molecular state $D^*\bar D^*$ of different
quantum numbers in this work. Using the effective interactions we
calculate the corresponding form factors for the decays
$Z_c(3900)\rightarrow h_c\pi$ and $Z_c(4020)\rightarrow h_c\pi$.  Our numerical results show
$\Gamma(Z_c(4020)(0^+)\rightarrow h_c\pi)$,
$\Gamma(Z_c(4020)(1^+)\rightarrow h_c\pi)$ and
$\Gamma(Z_c(4020)(2^+)\rightarrow h_c\pi)$  are indeed close to each
other. By the results one would
think that $Z_c(4020)$ behaves as a molecular state.

It is noticed that the resultant $\Gamma(Z_c(3900)\rightarrow h_c\pi)$ is only three
times smaller than $\Gamma(Z_c(4020)\rightarrow h_c\pi)$ for
various values of $\Lambda$ and $\beta$.

Considering the total width, even though the branching ratio of
$\Gamma(Z_c(3900)\rightarrow h_c\pi)$ is slightly small, we still
have a remarkable opportunity to observe $Z_c(3900)$ in this
channel. If  $Z_c(3900)$ and $Z_c(4020)$ are $D\bar D^*$ and
$D^*\bar D^*$ molecular states, we should observe the $Z_c(3900)$
peak at the invariant mass spectrum of $e^+e^-\to h_c\pi$. No
doubt, since this portal has not been ``seen" at BES III data so
far, the reason may be attributed to the relatively small database
at present. Thus with more data accumulating to a reasonable
stack, the experimental exploration of $Z_c(3900)\rightarrow
h_c\pi$ will eventually reach a conclusion, namely a peak at 3900
MeV shows up or does not. Namely, it does appear, one can
celebrate the assumption that $Z_c(3900)$ is indeed a molecular
state of $D\bar D^*(D^*\bar D)$ to be valid, or at least it
possess a large fraction of molecular state. By contrary, if there
is still no the signal of $Z_c(3900)$ to be observed at $h_c\pi$
invariant mass spectrum, the  the proposal that $Z_c(3900)$ were a
$D\bar D^*$ molecular state would not be favored or ruled out.

Even though in our calculation the coupling constant $g_{_{h_c
D^*D^*}}$ is  not well determined, so that the estimated widths
are not precise. However the ratio  $\Gamma(Z_c(3900)\to
h_c\pi)/\Gamma(Z_c(4020)\to h_c\pi)$ does not depend on the
coupling. Therefore, our scheme for judging whether $Z_c(3900)$ is
a molecular state is still working. A relevant question arises: what
is the inner structure of $Z_c(3900)$ if it is not a molecule? In
Ref.\cite{Dias:2013fza} the authors study some strong decays of
$Z_c(3900)$ by assuming it to be a tetraquark with the QCD sum rules, but
unfortunately the channel of $Z_c(3900)\rightarrow h_c\pi$ was not
discussed in their work. In our next work we will explore some strong decays of
$Z_c(3900)$ as a tetraquark especially including
$Z_c(3900)\rightarrow h_c\pi$ in the light front model, and will
show the partial width of this channel should indeed be small.

Since $Z_c(3900)$ was found from the final state $J/\psi\pi$, it
is natural to suggest that one should detect if $Z_c(4020)$ shows
up in the invariant mass spectrum of $J/\psi\pi$. Postulating both
$Z_c(3900)$ and $Z_c(4020)$ to be molecular states  we find
 $\Gamma(Z_c(4020)\rightarrow J/\psi\pi)$ is five times larger than $\Gamma(Z_c(3900)\rightarrow
 J/\psi\pi)$\cite{Ke:2013gia}. Thus we suggest our experimental colleagues to adjust the
center-mass-energy to produce a larger database for $Z_c(4020)$ to measure the corresponding decay rate.
It will be an ideal scheme to determine the identity of both $Z_c(3900)$ and $Z_c(4020)$.

Moreover, at the invariant mass spectrum of $D^*\bar{D^*}$,
another resonance $Z_c(4025)$ was measured with a mass of
$(4026.3\pm 2.6\pm3.7)$MeV and width $(24.8\pm 5.6\pm
7.7)$MeV\cite{Ablikim:2013emm}. Its peak heavily overlaps with
that of $Z_c(4020)$, and the deviation is within 1.5$\sigma$,
therefore it seems that $Z_c(4020)$ and $Z_c(4025)$ might be
degenerate, even more, they are the same state, but the
measurement errors cause a misidentification. Thus in the future
work it is our task to identify them as two different resonances
whose masses are close or just degenerate states or the same one.

\section*{Acknowledgement}
This work is supported by the National Natural Science Foundation
of China (NNSFC) under the contract No. 11375128 and 11135009.

\appendix
\section{the vertex function of molecular state}

Supposing $Z_c(3900)$ and $Z_c(4030)$ are  molecular states which
consists of $D$ and $\bar {D^*}$ and $D^{*}$ and $\bar {D^*}$
respectively. The wavefunction of a molecular state with total
spin $J$ and momentum $P$ is\cite{Ke:2013gia}
\begin{eqnarray}\label{eq:lfbaryon}
 |X(P,J,J_z)\rangle&=&\int\{d^3\tilde p_1\}\{d^3\tilde p_2\} \,
  2(2\pi)^3\delta^3(\tilde{P}-\tilde{p_1}-\tilde{p_2}) \nonumber\\
 &&\times\sum_{\lambda_1}\Psi^{SS_z}(\tilde{p}_1,\tilde{p}_2,\lambda_1,\lambda_2)
  \mathcal{F}\left|\right.
  D^{(*)}(p_1,\lambda_1) \bar D^*(p_2,\lambda_2)\ra.
\end{eqnarray}
For $0^+$ molecular state of $D^{*}\bar D^*$
\begin{eqnarray}
 \Psi^{SS_z}(\tilde{p}_1,\tilde{p}_2,\lambda_1,\lambda_2)&&=
 C_0 \varphi(x,p_{\perp})
\epsilon_{1}(\lambda_1)\cdot
 \epsilon_{2}(\lambda_2)\nonumber\\&&
=h_{C_0}' \epsilon_{1}(\lambda_1)\cdot
 \epsilon_{2}(\lambda_2),
\end{eqnarray}
for $1^+$ molecular state of $D^{*}\bar D^*$
\begin{eqnarray}
 \Psi^{SS_z}(\tilde{p}_1,\tilde{p}_2,\lambda_1,\lambda_2)&&=
 C_1 \varphi(x,p_{\perp})
\varepsilon^{\mu\nu\alpha\beta} \epsilon_{1\mu}(\lambda_1)
 \epsilon_{2\nu}(\lambda_2)\epsilon_{\alpha}(J_z)P_\beta\nonumber\\&&
=h_{C_1}' \varepsilon^{\mu\nu\alpha\beta}
\epsilon_{1\mu}(\lambda_1)
 \epsilon_{2\nu}(\lambda_2)\epsilon_{\alpha}(J_z)P_\beta,
\end{eqnarray}

for $2^+$ molecular state of $D^{*}\bar D^*$
\begin{eqnarray}
 \Psi^{SS_z}(\tilde{p}_1,\tilde{p}_2,\lambda_1,\lambda_2)&&=
 C_2 \varphi(x,p_{\perp})
\epsilon_{1\mu}(\lambda_1)
 \epsilon_{2\nu}(\lambda_2)\epsilon^{\mu\nu}(J_z)\nonumber\\&&
=h_{C_2}' \epsilon_{1\mu}(\lambda_1)
 \epsilon_{2\nu}(\lambda_2)\epsilon^{\mu\nu}(J_z),
\end{eqnarray}

and for $1^+$ molecular state of $D\bar D^*$
\begin{eqnarray}
 \Psi^{SS_z}(\tilde{p}_1,\tilde{p}_2,\lambda_1,\lambda_2)&&=
 C_{01(10)}\varphi(x,p_{\perp})
\epsilon_{1\mu}(\lambda_1)
 \cdot\epsilon_{\alpha}(J_z)\nonumber\\&&
=h_{C_{01(10)}}' \epsilon_{1\mu}(\lambda_1)
 \cdot\epsilon_{\alpha}(J_z),
\end{eqnarray}

{where $C_{01}, C_{10}, C_{0}, C_{1}$ and $C_{2}$ are the
normalization constants which can be fixed by normalizing the
state\cite{Cheng:2003sm}}
 \begin{eqnarray}\label{A6}
\langle X(P',J',J'_z)
|X(P,J,J_z)\rangle=2(2\pi)^3P^+\delta^3(\tilde{P}'-\tilde{P})\delta_{JJ'}\delta_{J_ZJ_{Z'}},
\end{eqnarray}
{and let the normailization
 $
\int\frac{dxd^2p_\perp}{2(2\pi)^3}\varphi'^*_{L',L'_Z}(x,p_\perp)\varphi_{L,L_Z}(x,p_\perp)=\delta_{_{L,L'}}\delta_{_{L_Z,L'_Z}}
$ hold.}

{For example $C_0$ is fixed by calculating Eq. (\ref{A6}) with the
$0^+$ state}
 \begin{eqnarray}\int\frac{dxd^2p_\perp}{2(2\pi)^3}C_0^2\epsilon_{1}^*(\lambda_1)\cdot
 \epsilon_{2}^*(\lambda_2)\epsilon_{1}(\lambda_1)\cdot
 \epsilon_{2}(\lambda_2)\varphi^*(x,p_\perp)\varphi(x,p_\perp)=1,
 \end{eqnarray}
{ then $C_0=\frac{2 {m_1} {m_2}}{\sqrt{{M_0}^4-2 {M_0}^2
({m_1}^2+{m_2}^2)+{m_1}^4+10 {m_1}^2 {m_2}^2+{m_2}^4}}$. It is
noted that $P^2=M_0^2$, $p1\cdot P=e_1M_0$ and $p2\cdot P=e_2M_0$
are used as discussed in Ref.\cite{Cheng:2003sm}. }

{Similarly one can obtain}
\begin{eqnarray*}
&&C_{01}=\frac{\sqrt{3}m_1}{\sqrt{e_1^2+2m_1^2}},\,\,\,\,\,\,\,\,\,\,\,
C_{10}=\frac{\sqrt{3}m_2}{\sqrt{e_2^2+2m_2^2}}, \nonumber\\&&
C_1=\frac{2\sqrt{3} {m_1} {m_2}}{\sqrt{{M}^2 [4 {e_1}^2 {m_2}^2-4
{e_1} {e_2} (-{M_0}^2+{m_1}^2+{m_2}^2)+4 {e_2}^2
   {m_1}^2+10 {m_1}^2 {m_2}^2-C_A]}}
, \nonumber\\&&C_2=\frac{\sqrt{120} {m_1} {m_2}}{\sqrt{4 {e_1}^2
(4 {e_2}^2+7 {m_2}^2)+4 {e_1} {e_2} (-{M_0}^2+{m_1}^2+{m_2}^2)+28
{e_2}^2
   {m_1}^2+54 {m_1}^2
   {m_2}^2+C_A}}, \nonumber\\&&C_A={M_0}^4-2 {M_0}^2
   ({m_1}^2+{m_2}^2)+{m_1}^4+{m_2}^4.
\end{eqnarray*}
and
$\varphi=4(\frac{\pi}{\beta^2})^{3/4}\frac{e_1e_2}{x_1x_2M_0}{\rm
exp}(\frac{-\mathbf{p}^2}{2\beta^2})$.

{ All other notations can be found in Ref}.\cite{Ke:2007tg}.
\section{the effective vertices}
 the effective vertices  can be found in
\cite{Haglin:1999xs,Oh:2000qr,Lin:1999ad,Deandrea:2003pv,Meng:2007cx},
\begin{eqnarray}\label{lagrangian_piDD}
 &&\mathcal L_{\pi DD^*}=ig_{_{\pi DD^*}}(D^{*\mu}\partial_\mu\pi\bar D - \partial^\mu D  \pi\bar D^{*}_\mu+h.c.)
 ,\\
 &&\mathcal L_{\pi D^*D^*}=-g_{_{\pi D^*D^*}}\varepsilon^{\mu\nu\alpha\beta}\partial_\mu\bar D^*_\nu\pi\partial_\alpha
 D^{*}_\beta
 ,\\
 &&\mathcal L_{h_c D^*D^*}=-ig_{_{h_c D^*D^*}}\varepsilon^{\mu\nu\alpha\beta}\partial_\mu {h_c}_{\nu} D^*_\alpha
\bar D^*_{\beta},\\
 &&\mathcal L_{h_c DD^*}=g_{_{h_c DD^*}}{h_c}^{\nu} D
\bar D^{*}_\nu.
 \end{eqnarray}


\begin{thebibliography}{99}
\bibitem{Ablikim:2013wzq}
  M.~Ablikim {\it et al.}  [BESIII Collaboration],
   Phys.\ Rev.\ Lett.\  {\bf 111}, 242001 (2013)  [arXiv:1309.1896 [hep-ex]].

\bibitem{Ablikim:2014dxl}
  M.~Ablikim {\it et al.} [BESIII Collaboration],
  Phys.\ Rev.\ Lett.\  {\bf 113}, 212002 (2014)  doi:10.1103/PhysRevLett.113.212002  [arXiv:1409.6577 [hep-ex]].  



\bibitem{Ablikim:2013mio}
  M.~Ablikim {\it et al.}  [BESIII Collaboration],
   Phys.\ Rev.\ Lett.\  {\bf 110}, 252001 (2013)  [arXiv:1303.5949 [hep-ex]].


\bibitem{Liu:2013dau}
  Z.~Q.~Liu {\it et al.}  [Belle Collaboration],
  Phys.\ Rev.\ Lett.\  {\bf 110}, 252002 (2013)  [arXiv:1304.0121 [hep-ex]].  

\bibitem{Xiao:2013iha}
  T.~Xiao, S.~Dobbs, A.~Tomaradze and K.~K.~Seth,
   Phys.\ Lett.\ B {\bf 727}, 366 (2013)  [arXiv:1304.3036 [hep-ex]].  

\bibitem{Esposito:2014hsa}
  A.~Esposito, A.~L.~Guerrieri and A.~Pilloni,
  Phys.\ Lett.\ B {\bf 746}, 194 (2015).

\bibitem{Li:2013xia}
  G.~Li,
   Eur.\ Phys.\ J.\ C {\bf 73}, 2621 (2013)   [arXiv:1304.4458 [hep-ph]].  

\bibitem{Deng:2014gqa}
  C.~Deng, J.~Ping and F.~Wang,
   Phys.\ Rev.\ D {\bf 90}, 054009 (2014)  doi:10.1103/PhysRevD.90.054009  [arXiv:1402.0777 [hep-ph]].  

\bibitem{Wang:2014gwa}
  Z.~G.~Wang,
   Eur.\ Phys.\ J.\ C {\bf 74}, 2963 (2014)  doi:10.1140/epjc/s10052-014-2963-7  [arXiv:1403.0810 [hep-ph]].  




\bibitem{Wang:2013cya}
  Q.~Wang, C.~Hanhart and Q.~Zhao,
   Phys.\ Rev.\ Lett.\  {\bf 111}, 132003 (2013)  [arXiv:1303.6355 [hep-ph]].  

\bibitem{Wilbring:2013cha}
  E.~Wilbring, H.-W.~Hammer and U.-G.~Mei$\beta$ner,
  Phys.\ Lett.\ B {\bf 726}, 326 (2013)  doi:10.1016/j.physletb.2013.08.059  [arXiv:1304.2882 [hep-ph]].  



\bibitem{Cui:2013yva}
  C.~Y.~Cui, Y.~L.~Liu, W.~B.~Chen and M.~Q.~Huang,
  J.\ Phys.\ G {\bf 41}, 075003 (2014)  [arXiv:1304.1850 [hep-ph]].  

\bibitem{Zhang:2013aoa}
  J.~R.~Zhang,
  Phys.\ Rev.\ D {\bf 87}, 116004 (2013)   [arXiv:1304.5748 [hep-ph]].  



\bibitem{Voloshin:2013dpa}
  M.~B.~Voloshin,
  Phys.\ Rev.\ D {\bf 87}, 091501 (2013)   [arXiv:1304.0380 [hep-ph]].

\bibitem{Liu:2013vfa}
  X.~H.~Liu and G.~Li,
   Phys.\ Rev.\ D {\bf 88}, 014013 (2013) [arXiv:1306.1384 [hep-ph]].  


\bibitem{Ke:2013gia}
  H.~W.~Ke, Z.~T.~Wei and X.~Q.~Li,
  Eur.\ Phys.\ J.\ C {\bf 73}, 2561 (2013)  [arXiv:1307.2414 [hep-ph]].  




\bibitem{Jaus} W. Jaus, Phys. Rev.  D {\bf 41}, 3394 (1990);
  D {\bf 44}, 2851 (1991);  W.~Jaus,
  Phys.\ Rev.\  D {\bf 60}, 054026 (1999).

\bibitem{Ji:1992yf}
  C.~R.~Ji, P.~L.~Chung and S.~R.~Cotanch,
  Phys.\ Rev.\  D {\bf 45}, 4214 (1992).
\bibitem{Cheng:2004cc}
  H.~-Y.~Cheng, C.~-K.~Chua and C.~-W.~Hwang,
   Phys.\ Rev.\ D {\bf 70}, 034007 (2004)  [hep-ph/0403232].  


\bibitem{Ke:2007tg}
  H.~W.~Ke, X.~Q.~Li and Z.~T.~Wei,
  Phys.\ Rev.\  D {\bf 77}, 014020 (2008)
  [arXiv:0710.1927 [hep-ph]];
  Z.~T.~Wei, H.~W.~Ke and X.~Q.~Li,
  Phys.\ Rev.\  D {\bf 80}, 094016 (2009)
  [arXiv:0909.0100 [hep-ph]];
  H.~-W.~Ke, X.~-H.~Yuan, X.~-Q.~Li, Z.~-T.~Wei and Y.~-X.~Zhang,
   Phys.\ Rev.\ D {\bf 86}, 114005 (2012)  [arXiv:1207.3477 [hep-ph]].  


\bibitem{Ke:2009ed}
  H.~W.~Ke, X.~Q.~Li and Z.~T.~Wei,
  Phys.\ Rev.\  D {\bf 80}, 074030 (2009)
  [arXiv:0907.5465 [hep-ph]];
%
  H.~W.~Ke, X.~Q.~Li and Z.~T.~Wei,
  Eur.\ Phys.\ J.\  C {\bf 69}, 133 (2010)
  [arXiv:0912.4094 [hep-ph]];
  H.~W.~Ke, X.~H.~Yuan and X.~Q.~Li,
Int. J. Mod. Phys. A {\bf 26}, 4731 (2011),
  arXiv:1101.3407 [hep-ph];
  H.~W.~Ke and X.~Q.~Li,
  Eur.\ Phys.\ J.\  C {\bf 71}, 1776 (2011)
  [arXiv:1104.3996 [hep-ph]].




\bibitem{Cheng:1996if}
  H.~Y.~Cheng, C.~Y.~Cheung and C.~W.~Hwang,
  Phys.\ Rev.\  D {\bf 55}, 1559 (1997)
  [arXiv:hep-ph/9607332].

\bibitem{Li:2010bb}
  G.~Li, F.~l.~Shao and W.~Wang,
  Phys.\ Rev.\  D {\bf 82}, 094031 (2010)
  [arXiv:1008.3696 [hep-ph]].

\bibitem{Cheng:2003sm}
  H.~Y.~Cheng, C.~K.~Chua and C.~W.~Hwang,
  Phys.\ Rev.\  D {\bf 69}, 074025 (2004).


\bibitem{Hwang:2006cua}
  C.~W.~Hwang and Z.~T.~Wei,
  J.\ Phys.\ G {\bf 34}, 687 (2007);
%
  C.~D.~Lu, W.~Wang and Z.~T.~Wei,
  Phys.\ Rev.\  D {\bf 76}, 014013 (2007)
  [arXiv:hep-ph/0701265].


\bibitem{Choi:2007se}
  H.~M.~Choi,
  Phys.\ Rev.\  D {\bf 75}, 073016 (2007)
  [arXiv:hep-ph/0701263];

\bibitem{Ke:2013zs}
  H.~-W.~Ke, X.~-Q.~Li and Y.~-L.~Shi,
 Phys.\ Rev.\  D {\bf 87}, 054022 (2013)
  arXiv:1301.4014 [hep-ph]; 

  H.~W.~Ke, X.~Q.~Li, Z.~T.~Wei and X.~Liu,
  Phys.\ Rev.\  D {\bf 82}, 034023 (2010)
  [arXiv:1006.1091 [hep-ph]].



\bibitem{Haglin:1999xs}
  K.~L.~Haglin,
  Phys.\ Rev.\ C {\bf 61} (2000) 031902.

\bibitem{Oh:2000qr}
  Y.~-S.~Oh, T.~Song and S.~H.~Lee,
  Phys.\ Rev.\ C {\bf 63}, 034901 (2001)
  [nucl-th/0010064].

\bibitem{Lin:1999ad}
  Z.~-W.~Lin and C.~M.~Ko,
  Phys.\ Rev.\ C {\bf 62}, 034903 (2000).

\bibitem{Deandrea:2003pv}
 A.~Deandrea, G.~Nardulli and A.~D.~Polosa,
 Phys.\ Rev.\ D {\bf 68}, 034002 (2003)[hep-ph/0302273].

\bibitem{Meng:2007cx}
  C.~Meng and K.~-T.~Chao,
  Phys.\ Rev.\ D {\bf 75}, 114002 (2007)
  [hep-ph/0703205].



\bibitem{Yuan:2012zw}
  X.~-H.~Yuan, H.~-W.~Ke, X.~Liu and X.~-Q.~Li,
   Phys.\ Rev.\ D {\bf 87}, 014019 (2013)  [arXiv:1210.3686 [hep-ph]].  




\bibitem{PDG12}
  K.~A.~Olive {\it et al.}  [Particle Data Group Collaboration],
  Chin.\ Phys.\ C {\bf 38}, 090001 (2014).




\bibitem{Ke:2011jf}
  H.~W.~Ke and X.~Q.~Li,
  Phys.\ Rev.\  D {\bf 84}, 114026 (2011)
  [arXiv:1107.0443 [hep-ph]];


\bibitem{Dias:2013fza}
  J.~M.~Dias, F.~S.~Navarra, M.~Nielsen and C.~Zanetti,
  arXiv:1311.7591 [hep-ph].  

\bibitem{Ablikim:2013emm}
  M.~Ablikim {\it et al.} [BESIII Collaboration],
  Phys.\ Rev.\ Lett.\  {\bf 112}, 132001 (2014)
  [arXiv:1308.2760 [hep-ex]].

\end{thebibliography}
\end{document}